# Elsevier’s Pre-proof Policy Blocks Google Scholar Indexing

Shi-Shun Chen

***Abstract*—Google Scholar is a vital tool for engineering scholars, enabling efficient literature searches and facilitating academic dissemination. Elsevier, as one of the largest publishers of engineering journals, produces essential research that scholars rely on. The pre-proof policy, adopted by Elsevier for certain journals, allows articles to be published online in their accepted draft form before final proofreading and formatting. However, this study empirically demonstrates that the pre-proof publication policy hinders comprehensive indexing by Google Scholar. Articles published under this policy are only partially indexed—often limited to titles and abstracts—while crucial sections such as introductions, methods, results, discussions, conclusions, appendices, and data availability statements remain unsearchable. This problem has persisted for years, resulting in reduced visibility and accessibility of certain Elsevier articles. To improve academic dissemination, both Elsevier and Google Scholar must address this problem by modifying publishing policies or enhancing indexing practices. Additionally, this paper explores strategies that authors can use to mitigate the issue and ensure broader discoverability of their research.**

## I. Introduction

For a significant number of Elsevier journals, once an article is accepted, the accepted manuscript is directly published online in the journal's "Articles in Press" section. According to the definition by Elsevier, contents in "Articles in Press" are “Articles in press are accepted, peer reviewed articles that are not yet assigned to volumes/issues, but are citable using DOI.” In this paper, we will demonstrate through empirical research that this publishing policy significantly hinders their comprehensive indexing by Google Scholar.

First, our study presents cases of indexing failures, including indexing Failure on all contents and part contents, and compares them with articles published without the pre-proof policy to clarify how proper indexing should work. Then, we talk about the harmful influence of these problems. Finally, we conclude the paper with our findings.

Shi-Shun Chen is with the School of Reliability and Systems Engineering, Beihang University, Beijing 100191, China (e-mail: css1107@buaa.edu.cn).

## II. Indexing Failure Cases

### A. Indexing Failure on all contents

Here, the articles are published in 5 random selected journals by Elsevier using pre-proof publication policy: *Mechanical Systems and Signal Processing* (MSSP), *Reliability Engineering & System Safety* (RESS), *Computers & Industrial Engineering* (CIE), *Applied Mathematical Modelling* (AMM), and *Swarm and Evolutionary Computation* (SEC).

We conducted our search using Google Scholar's strategy, which requires quotation marks to enclose mandatory terms. For the articles mentioned in this subsection, indexing is limited to titles and abstracts, leaving out all other contents, including introductions, methods, results, discussions, conclusions, appendices, and data availability statements. Notably, these are only some examples, and the publication year ranges from 2021 to 2024.

1) *Mechanical Systems and Signal Processing*

Reference: [1]

Doi: https://doi.org/10.1016/j.ymssp.2023.110923

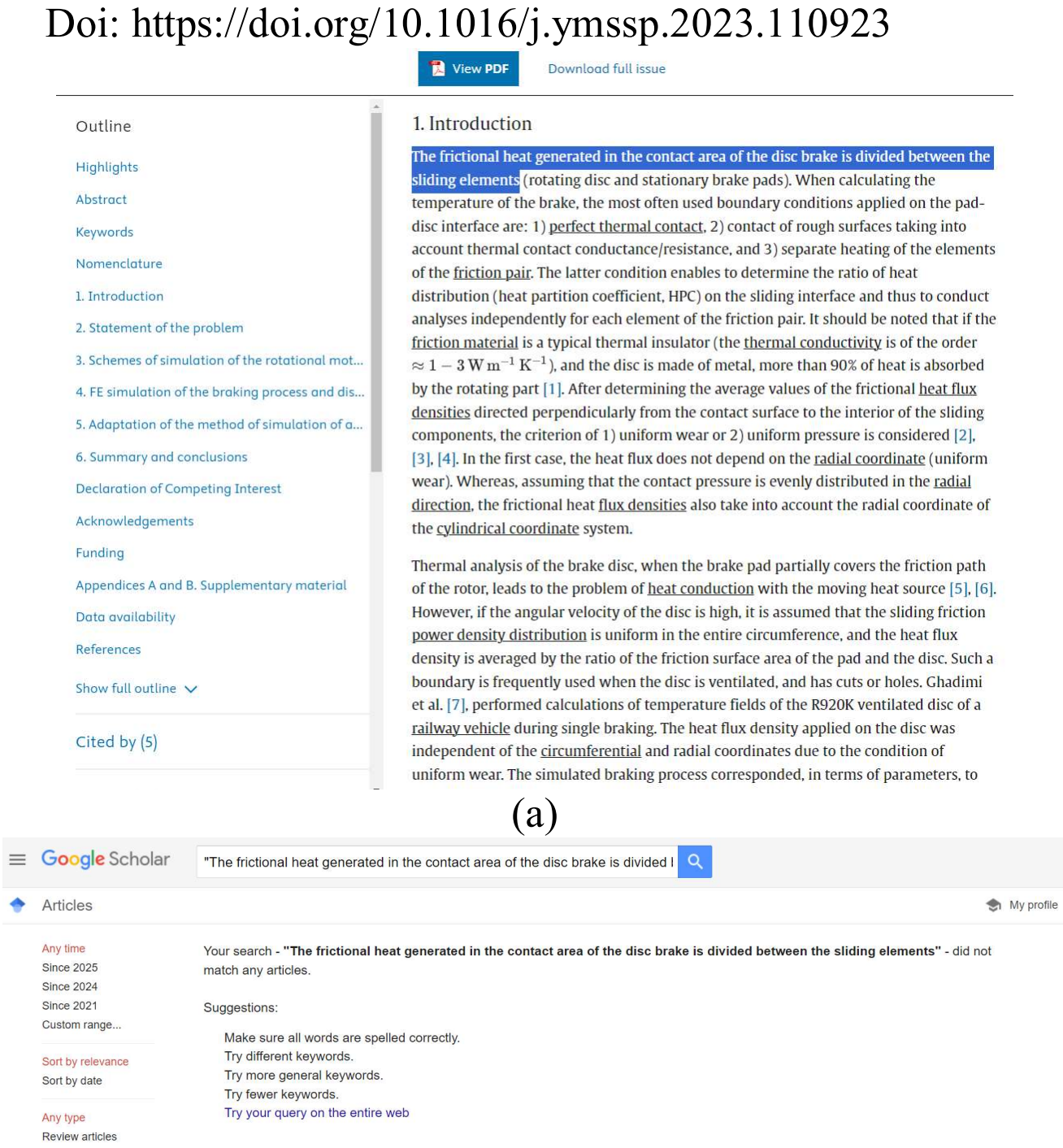

(a)

(b)

Fig. 1 Indexing Failure on all contents of [1] published in MSSP.

2) *Reliability Engineering & System Safety*

Reference: [2]

Doi: https://doi.org/10.1016/j.ress.2024.109964

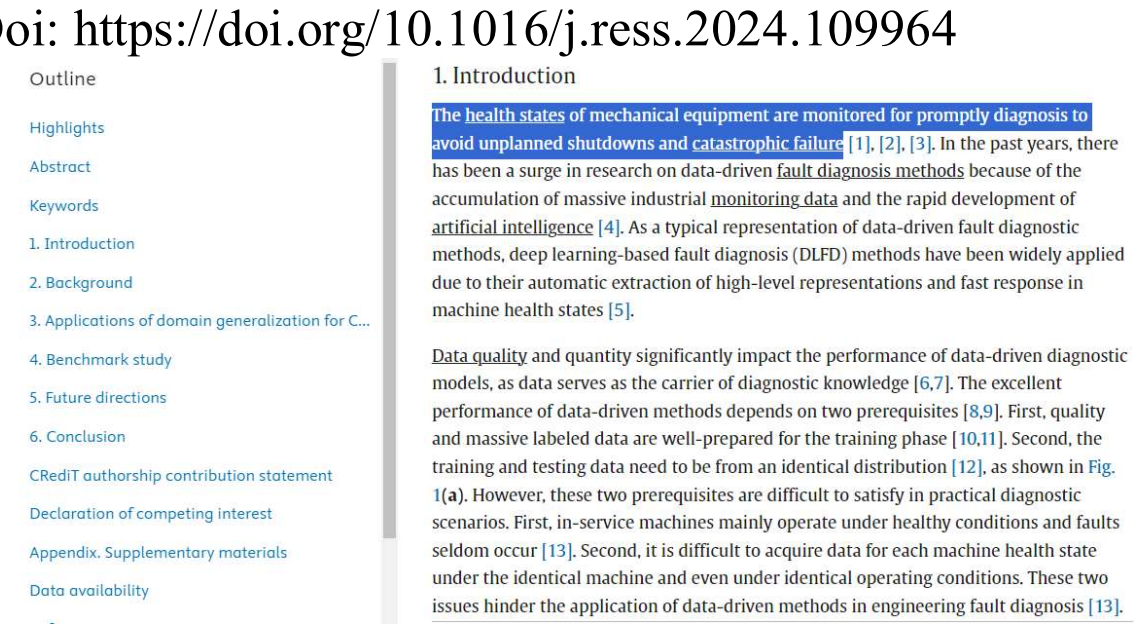

(a)

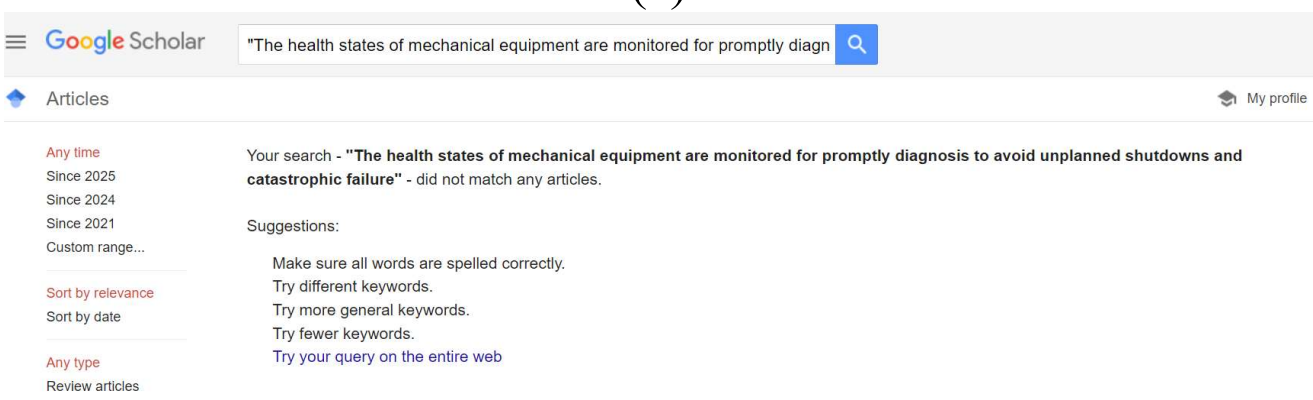

(b)

Fig. 2 Indexing Failure on all contents of [2] published in RESS.

3) *Computers & Industrial Engineering*

Reference: [3]

Doi: https://doi.org/10.1016/j.cie.2022.108551

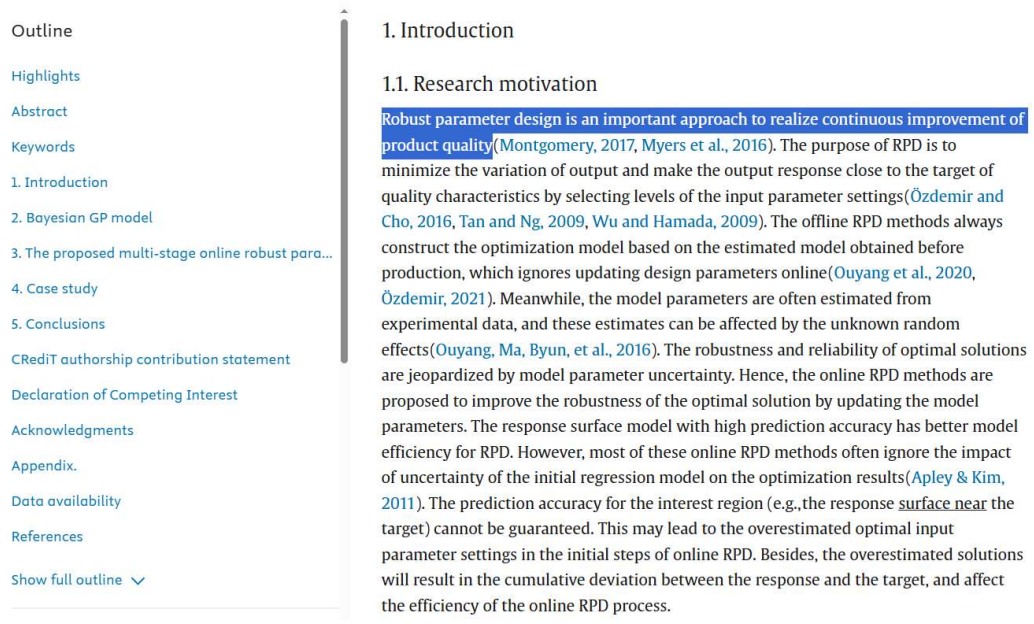

(a)

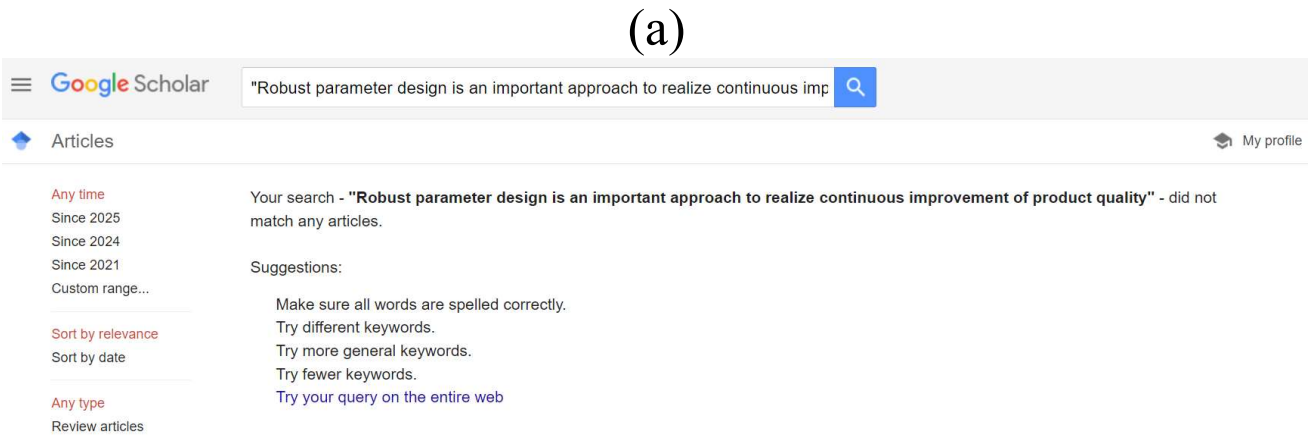

(b)

Fig. 3 Indexing Failure on all contents of [3] published in CIE.

4) *Applied Mathematical Modelling*

Reference: [4]

Doi: https://doi.org/10.1016/j.apm.2024.115788

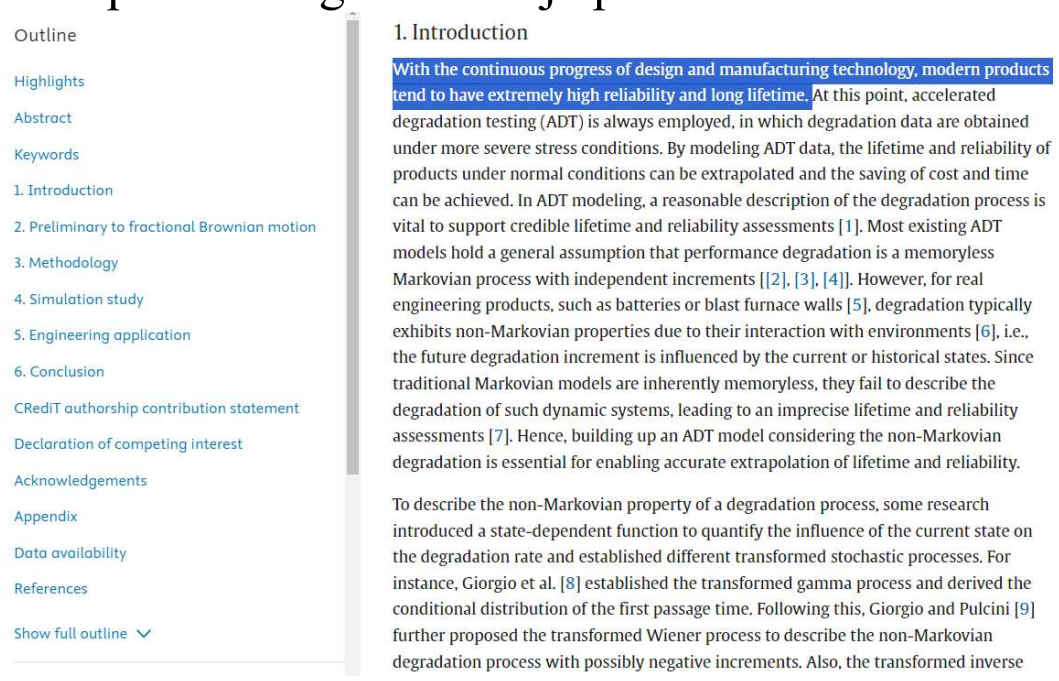

(a)

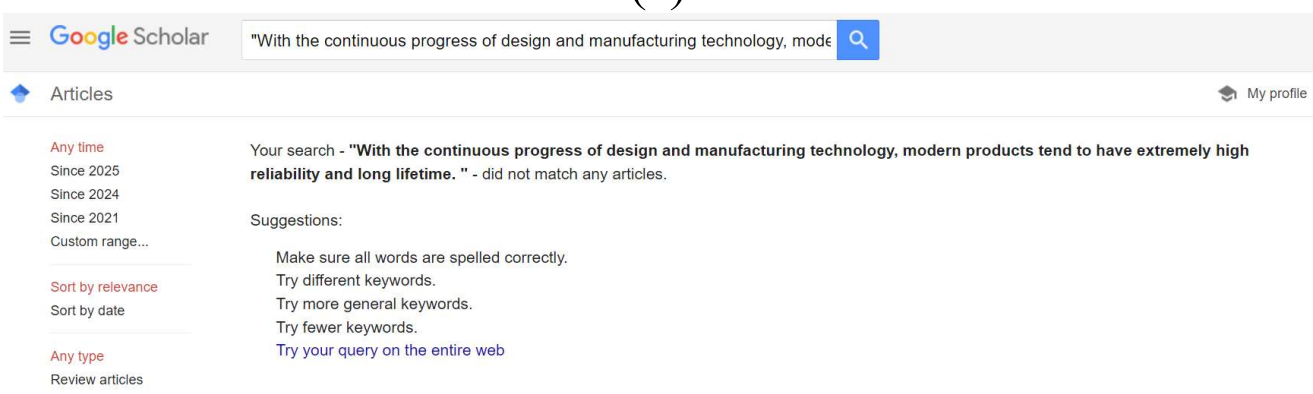

(b)

Fig. 4 Indexing Failure on all contents of [4] published in AMM.

5) *Swarm and Evolutionary Computation*

Reference: [5]

Doi: https://doi.org/10.1016/j.swevo.2021.100927

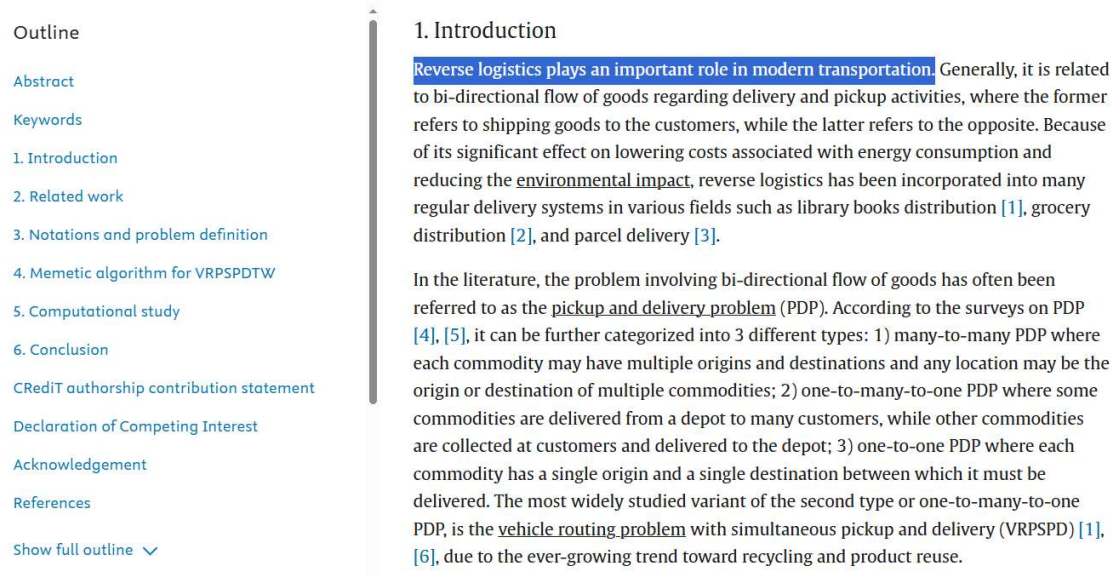

(a)

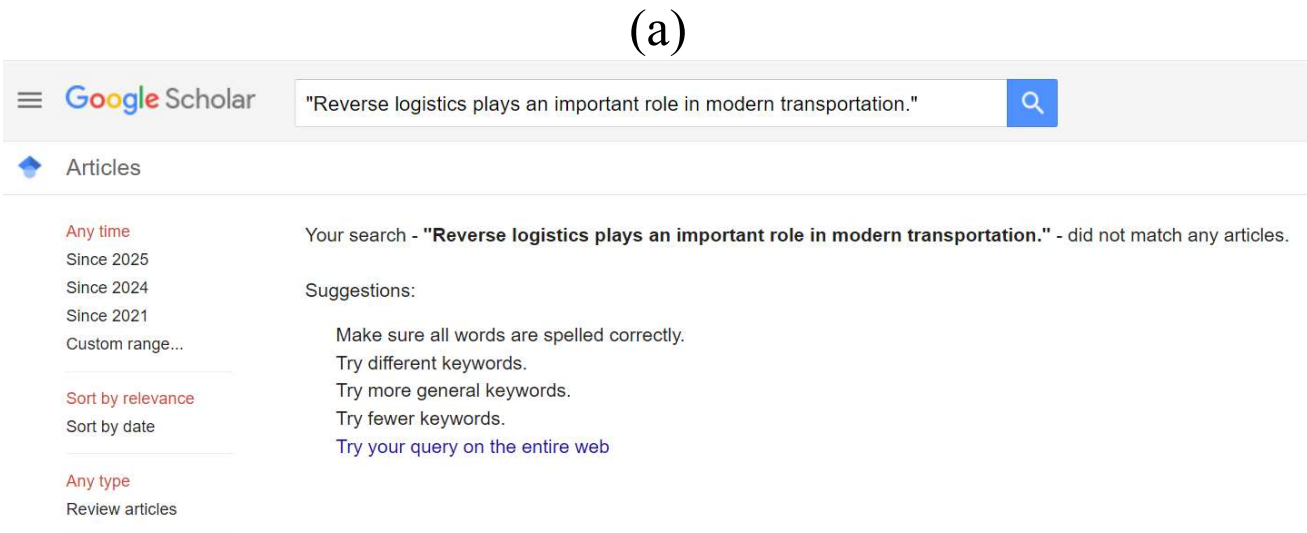

(b)

Fig. 5 Indexing Failure on all contents of [5] published in SEC.

## B. Indexing Failure on part contents

Actually, 99% of the articles published in journals by Elsevier using pre-proof publication policy have this problem. This can be easily checked by searching the content in the "conclusion" or "data availability" section on Google Scholar.

In this subsection, we want to talk about something interesting. We found that for all articles published under the pre-proof policy, their indexing is disrupted by a certain figure, table, paragraph or even punctuation, resulting in only the former portion being indexed, while the remain portion is not.

1) *Indexing disrupted by figure*

Reference: [6]

Doi: https://doi.org/10.1016/j.matcom.2024.10.012

It can be seen in Fig. 6 that using the words above Table 4 and those in Table 4 can search this article, whereas using the words in Fig. 5 and those after Fig. 5 can't search this article.

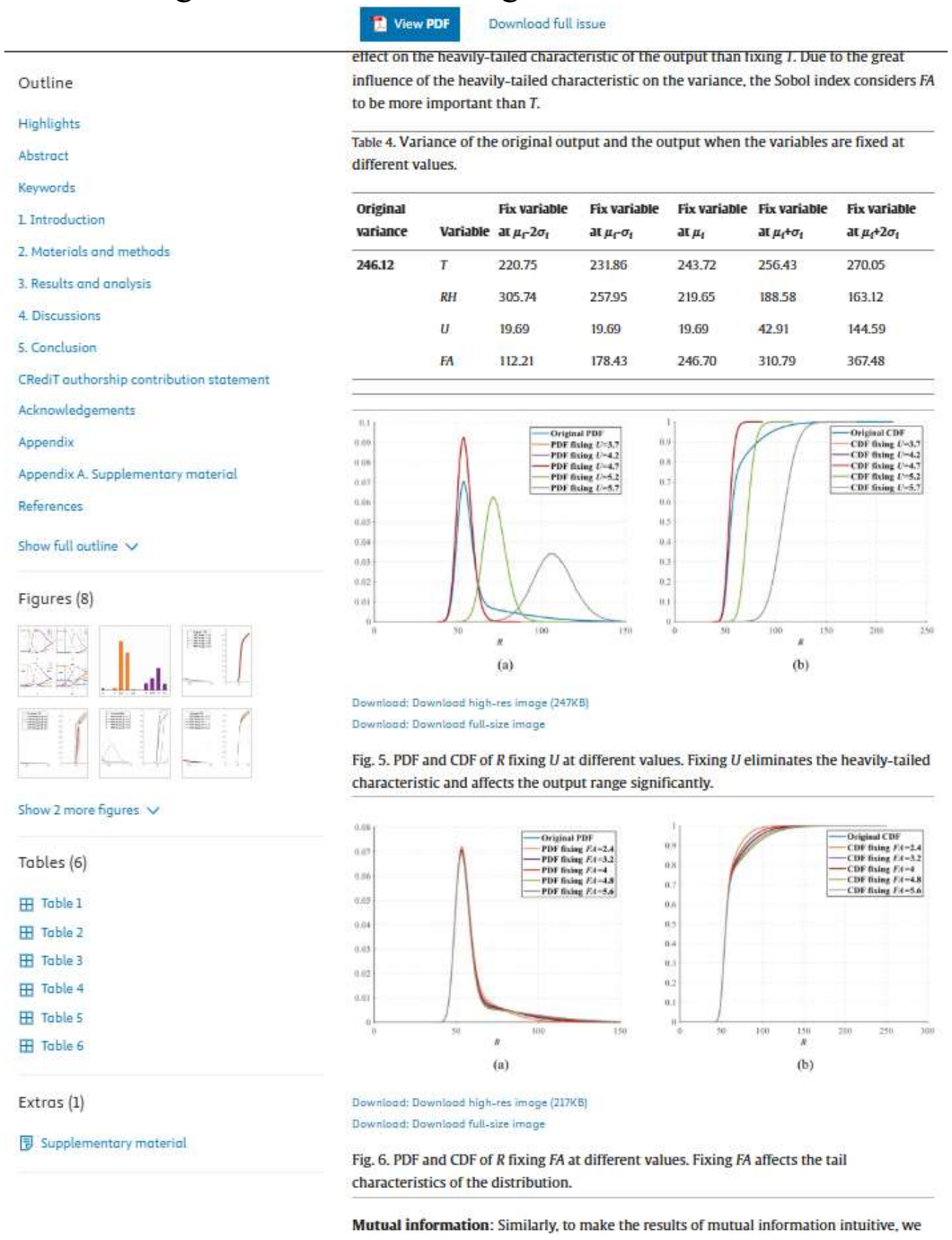

| Original variance | Variable | Fix variable at $\mu_i$-2$\sigma_i$ | Fix variable at $\mu_i$-$\sigma_i$ | Fix variable at $\mu_i$ | Fix variable at $\mu_i$+$\sigma_i$ | Fix variable at $\mu_i$+2$\sigma_i$ |
|---|---|---|---|---|---|---|
| 246.12 | T | 220.75 | 231.86 | 243.72 | 256.43 | 270.05 |
| | RH | 305.74 | 257.95 | 219.65 | 188.58 | 163.12 |
| | U | 19.69 | 19.69 | 19.69 | 42.91 | 144.59 |
| | FA | 112.21 | 178.43 | 246.70 | 310.79 | 367.48 |

(a)

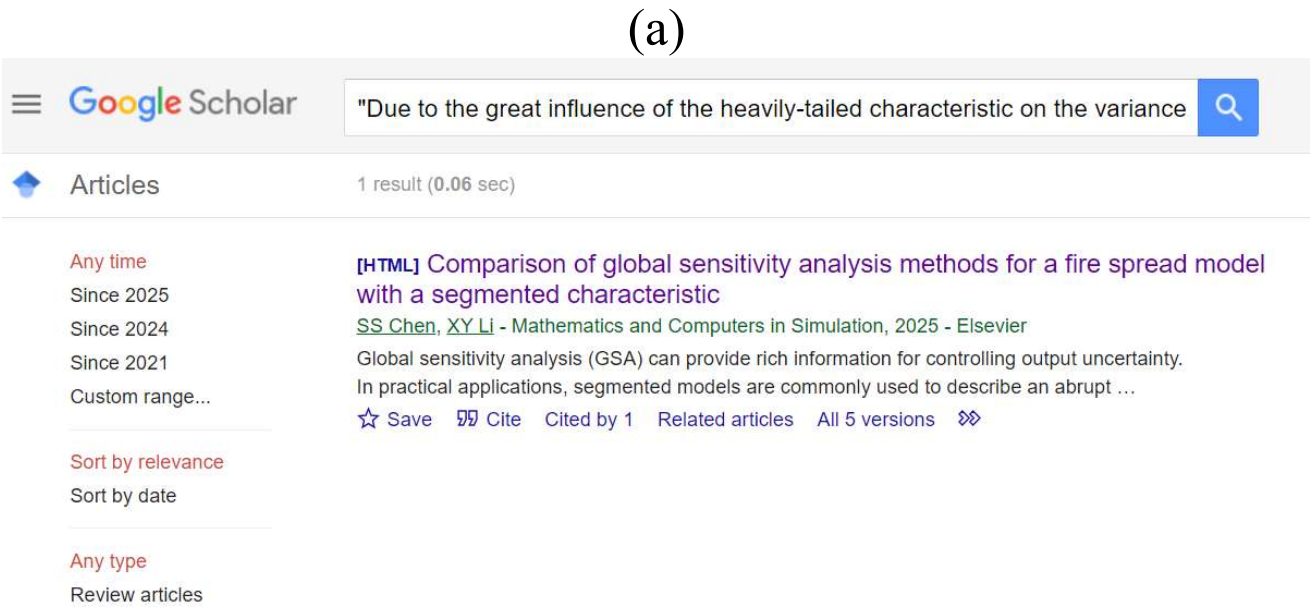

(b)

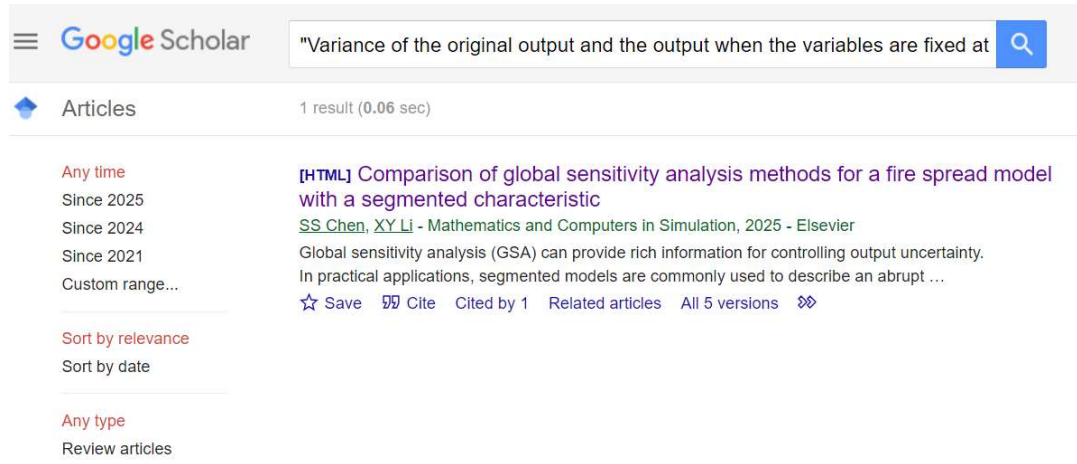

(c)

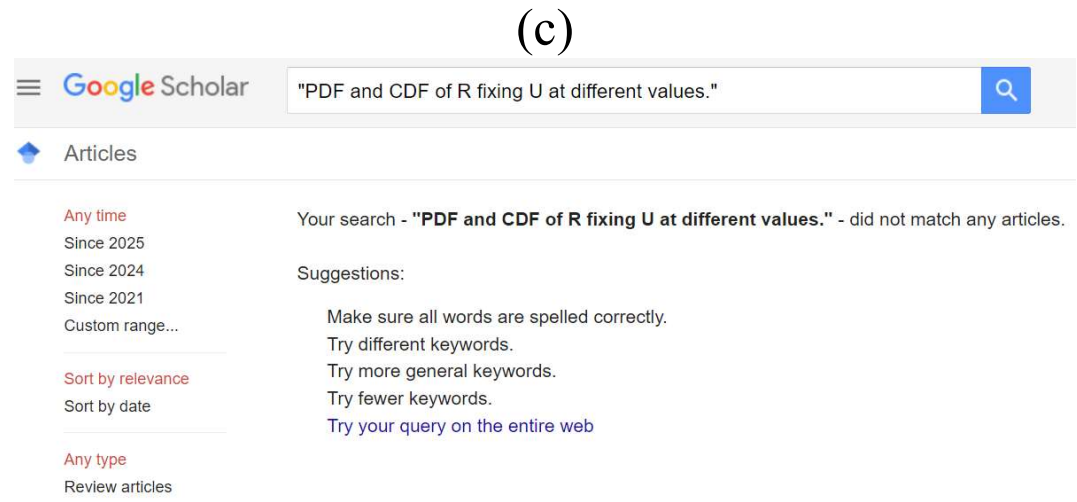

(d)

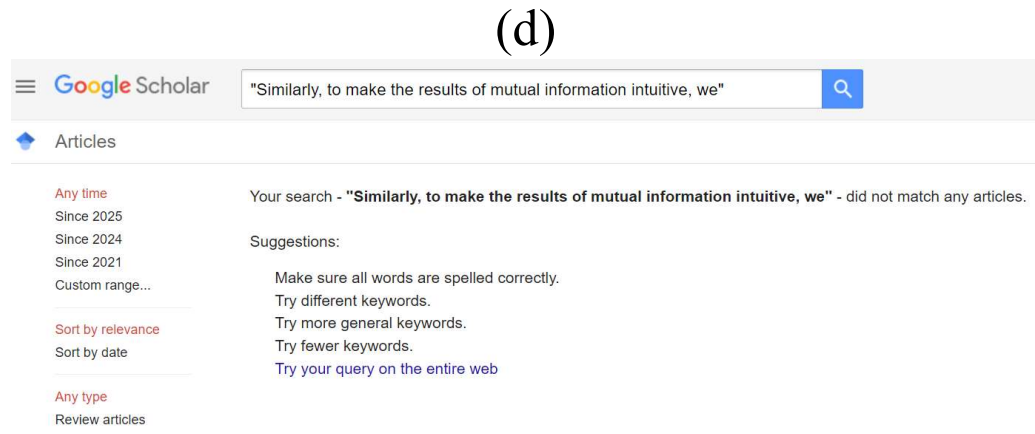

(e)

Fig. 6 Indexing disrupted by the figure published in *Mathematics and Computers in Simulation* [6].

2) *Indexing disrupted by table*

Reference: [7]

Doi: https://doi.org/10.1016/j.inffus.2020.06.007

It can be seen in Fig. 7 that using the words in Table 5 can search this article, whereas using the words in Table 6 and those after Table 6 can't search this article.

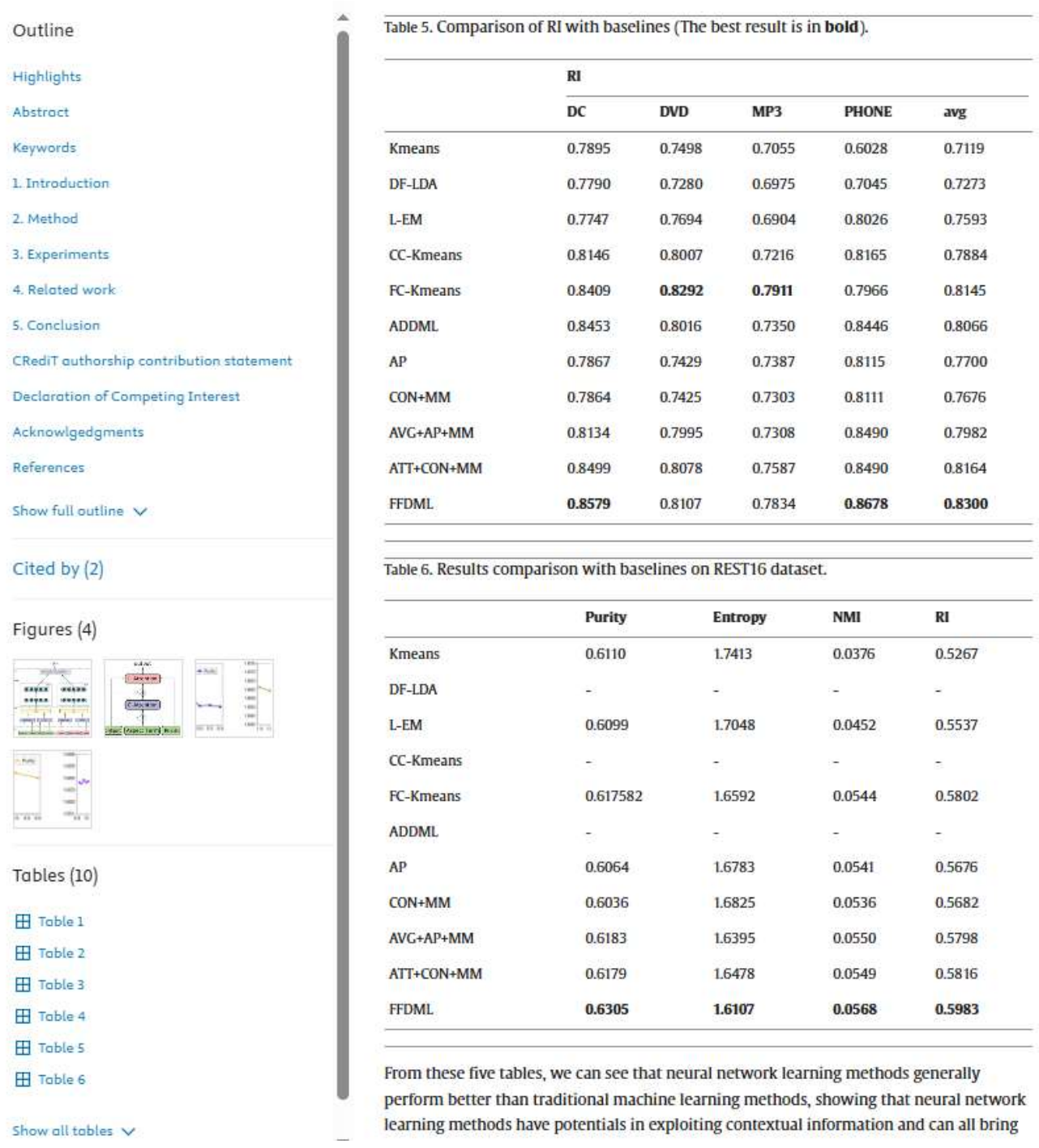

| | RI | | | | |
|---|---|---|---|---|---|
| | DC | DVD | MP3 | PHONE | avg |
| Kmeans | 0.7895 | 0.7498 | 0.7055 | 0.6028 | 0.7119 |
| DF-LDA | 0.7790 | 0.7280 | 0.6975 | 0.7045 | 0.7273 |
| L-EM | 0.7747 | 0.7694 | 0.6904 | 0.8026 | 0.7593 |
| CC-Kmeans | 0.8146 | 0.8007 | 0.7216 | 0.8165 | 0.7884 |
| FC-Kmeans | 0.8409 | **0.8292** | **0.7911** | 0.7966 | 0.8145 |
| ADDML | 0.8453 | 0.8016 | 0.7350 | 0.8446 | 0.8066 |
| AP | 0.7867 | 0.7429 | 0.7387 | 0.8115 | 0.7700 |
| CON+MM | 0.7864 | 0.7425 | 0.7303 | 0.8111 | 0.7676 |
| AVG+AP+MM | 0.8134 | 0.7995 | 0.7308 | 0.8490 | 0.7982 |
| ATT+CON+MM | 0.8499 | 0.8078 | 0.7587 | 0.8490 | 0.8164 |
| FFDML | **0.8579** | 0.8107 | 0.7834 | **0.8678** | **0.8300** |

| | Purity | Entropy | NMI | RI |
|---|---|---|---|---|
| Kmeans | 0.6110 | 1.7413 | 0.0376 | 0.5267 |
| DF-LDA | - | - | - | - |
| L-EM | 0.6099 | 1.7048 | 0.0452 | 0.5537 |
| CC-Kmeans | - | - | - | - |
| FC-Kmeans | 0.617582 | 1.6592 | 0.0544 | 0.5802 |
| ADDML | - | - | - | - |
| AP | 0.6064 | 1.6783 | 0.0541 | 0.5676 |
| CON+MM | 0.6036 | 1.6825 | 0.0536 | 0.5682 |
| AVG+AP+MM | 0.6183 | 1.6395 | 0.0550 | 0.5798 |
| ATT+CON+MM | 0.6179 | 1.6478 | 0.0549 | 0.5816 |
| FFDML | **0.6305** | **1.6107** | **0.0568** | **0.5983** |

(a)

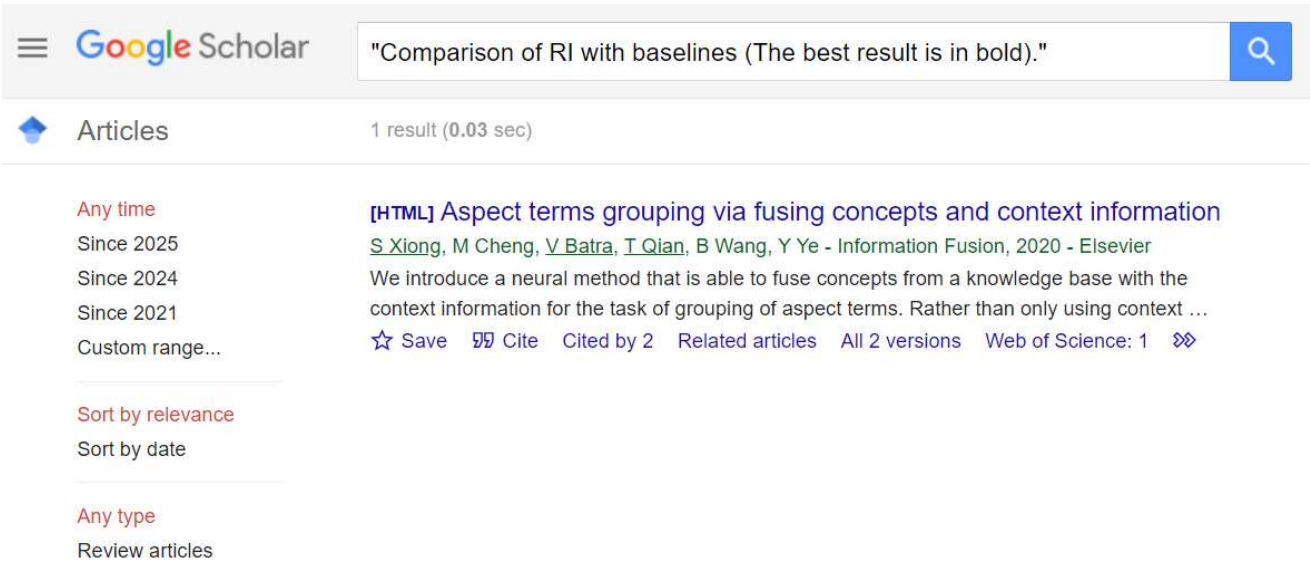

(b)

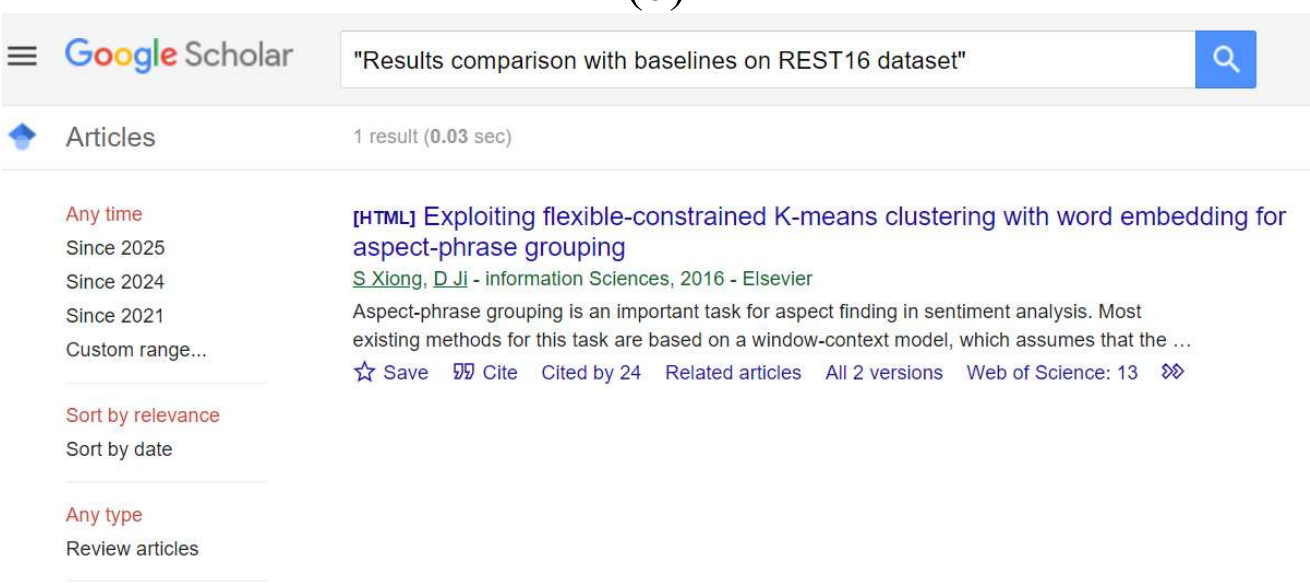

(c)

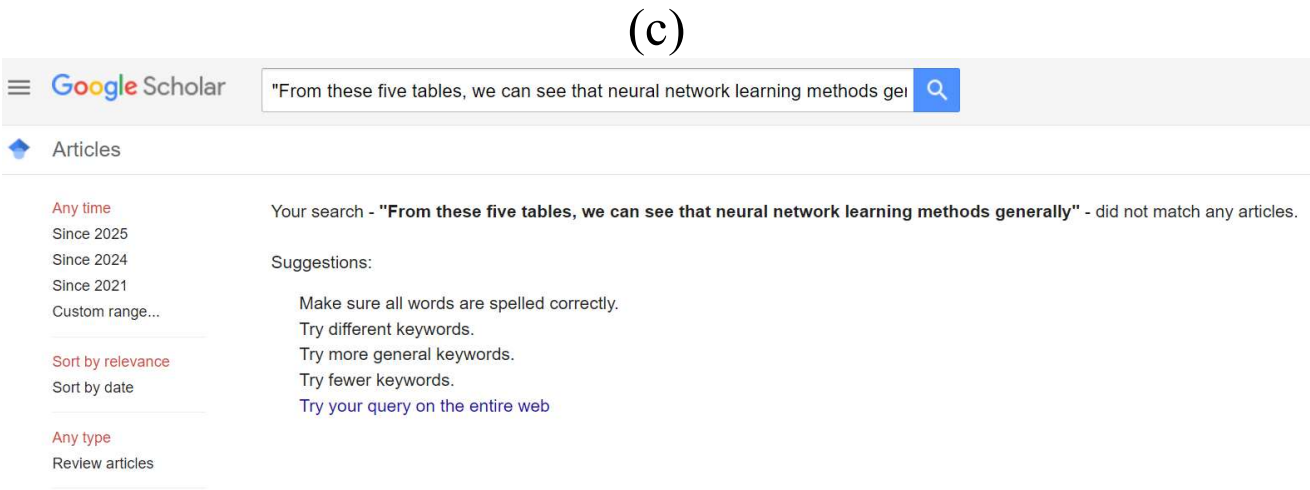

(d)

Fig. 7 Indexing disrupted by the paragraph published in *Information Fusion* [7].

### 3) *Indexing disrupted by paragraph*

Reference: [8]

Doi: https://doi.org/10.1016/j.ast.2024.109848

It can be seen in Fig. 8 that using the words in the first paragraph can search this article, whereas using the words in the second paragraph can't search this article.

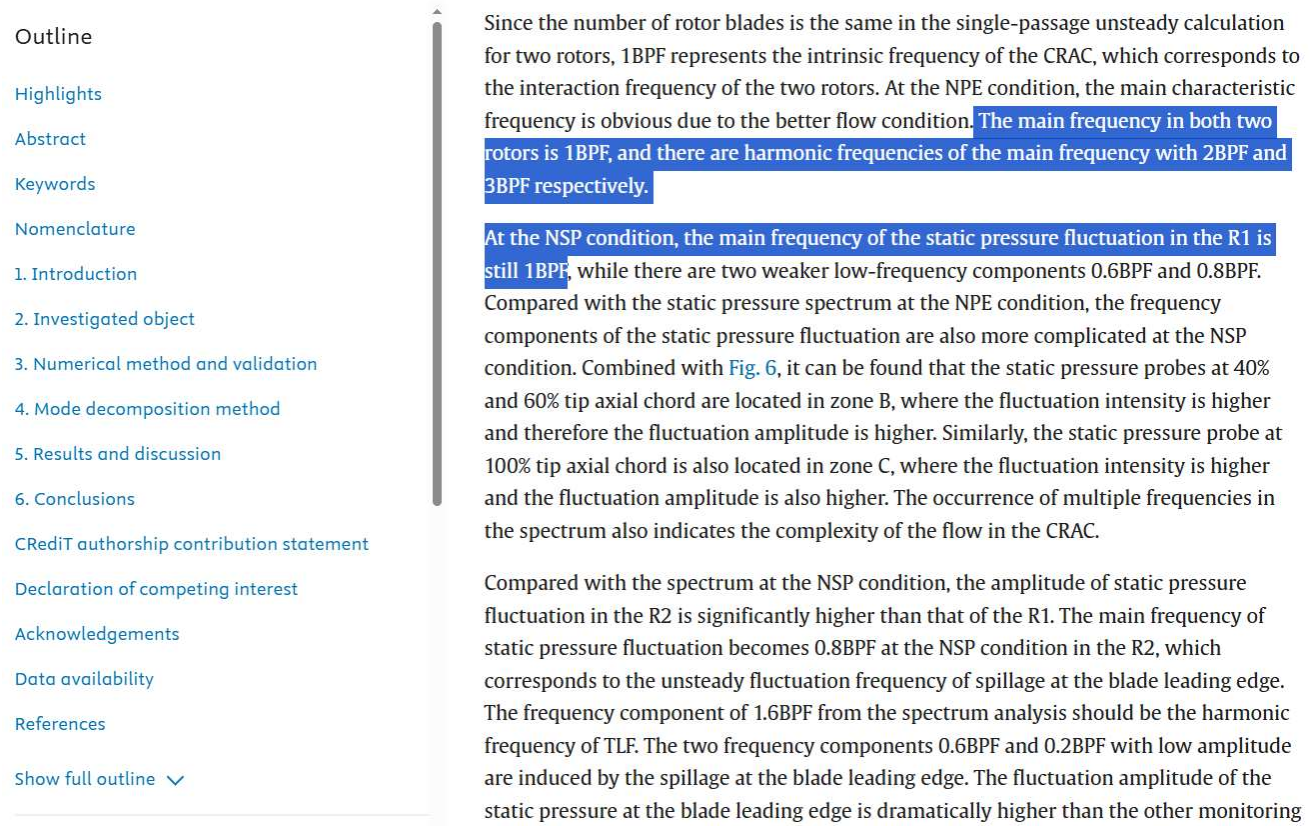

(a)

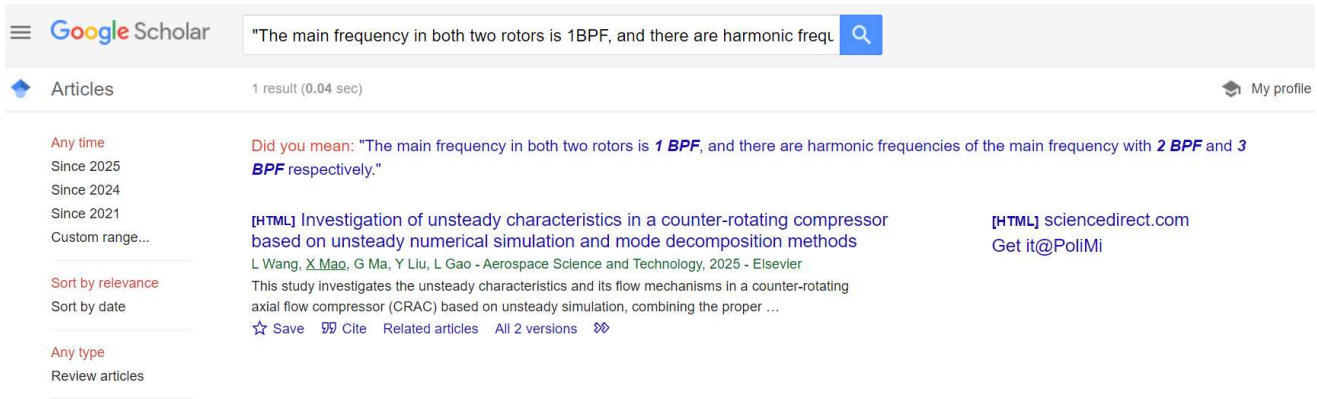

(b)

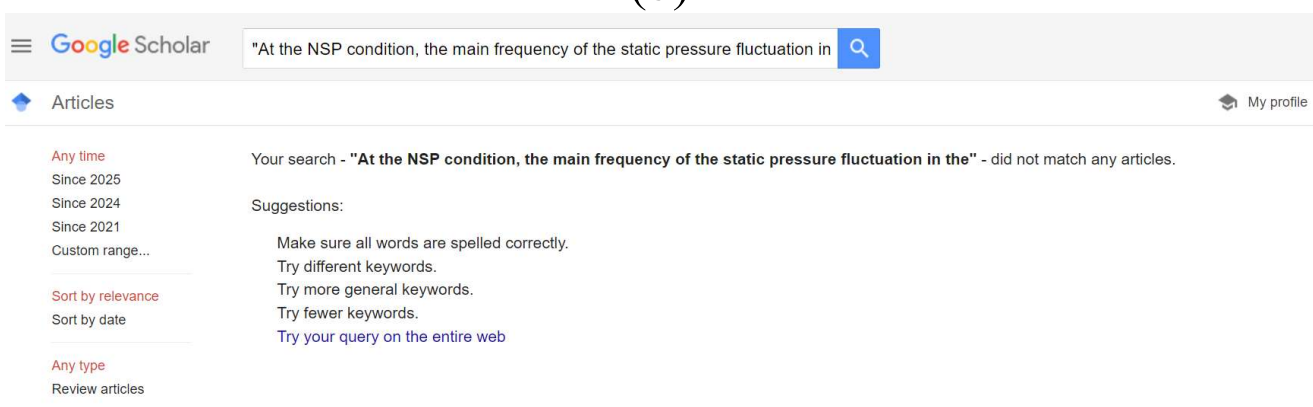

(c)

Fig. 8 Indexing disrupted by the paragraph published in *Aerospace Science and Technology* [8].

### 4) *Indexing disrupted by punctuation*

Reference: [9]

Doi: https://doi.org/10.1016/j.jii.2024.100709

It can be seen in Fig. 9 that using the words before the colon can search this article, whereas using the words after the colon can't search this article.

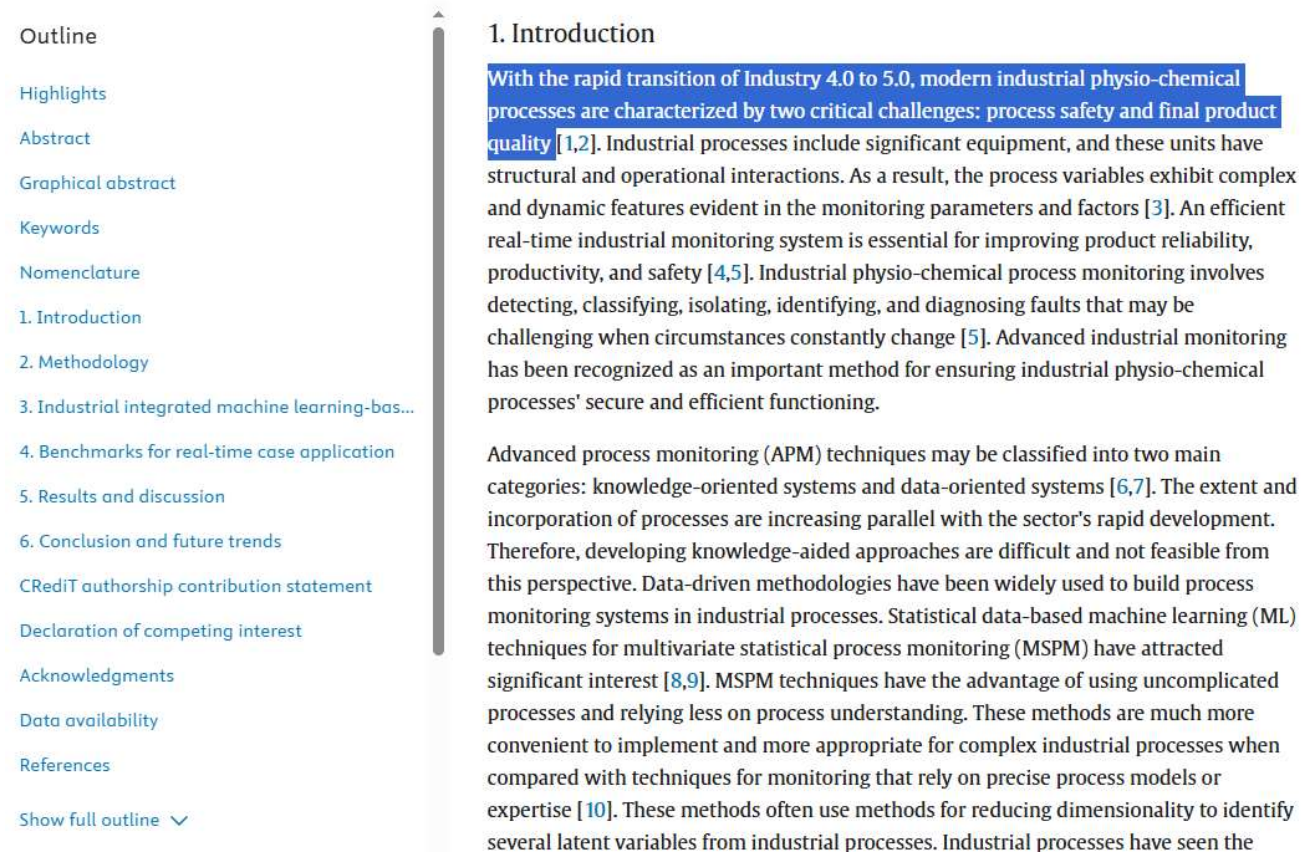

(a)

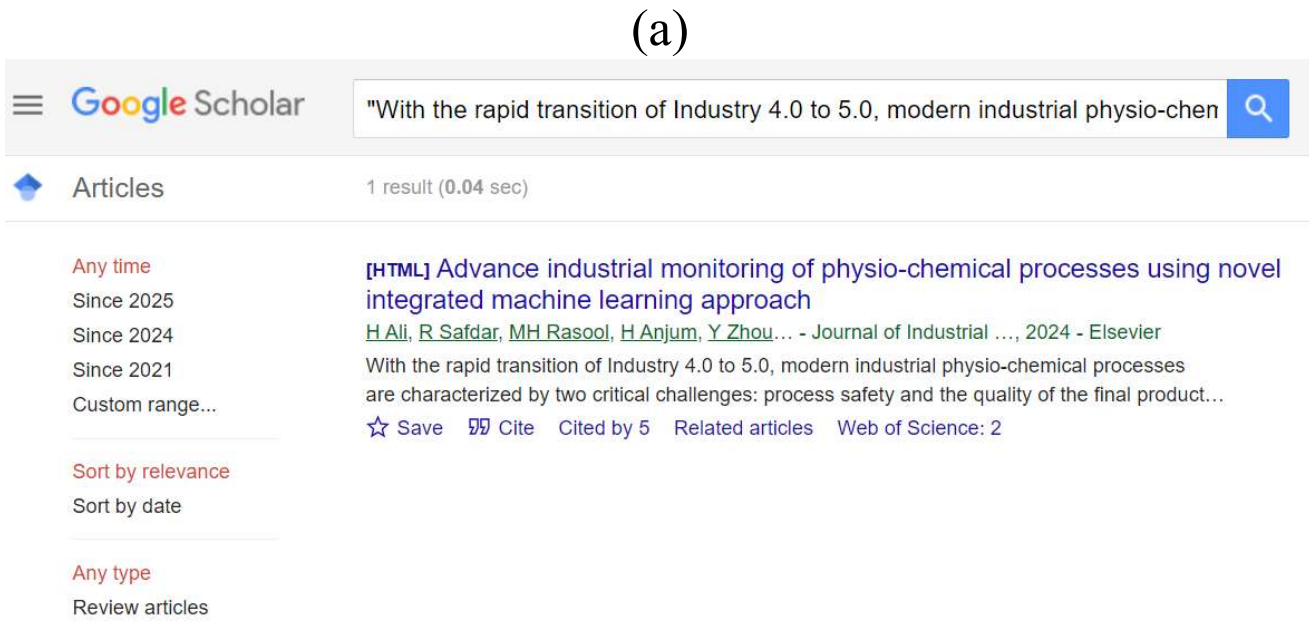

(b)

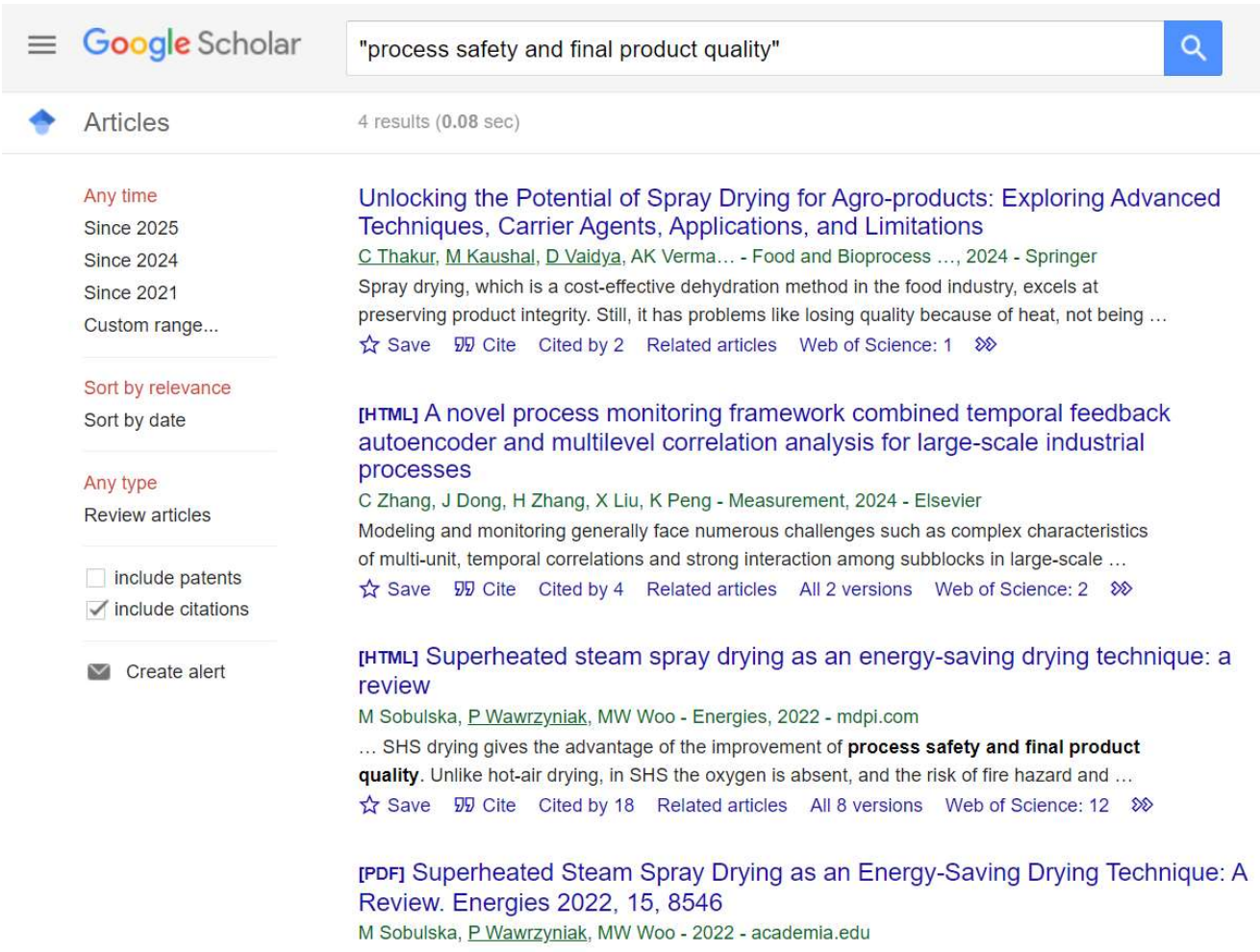

(c)

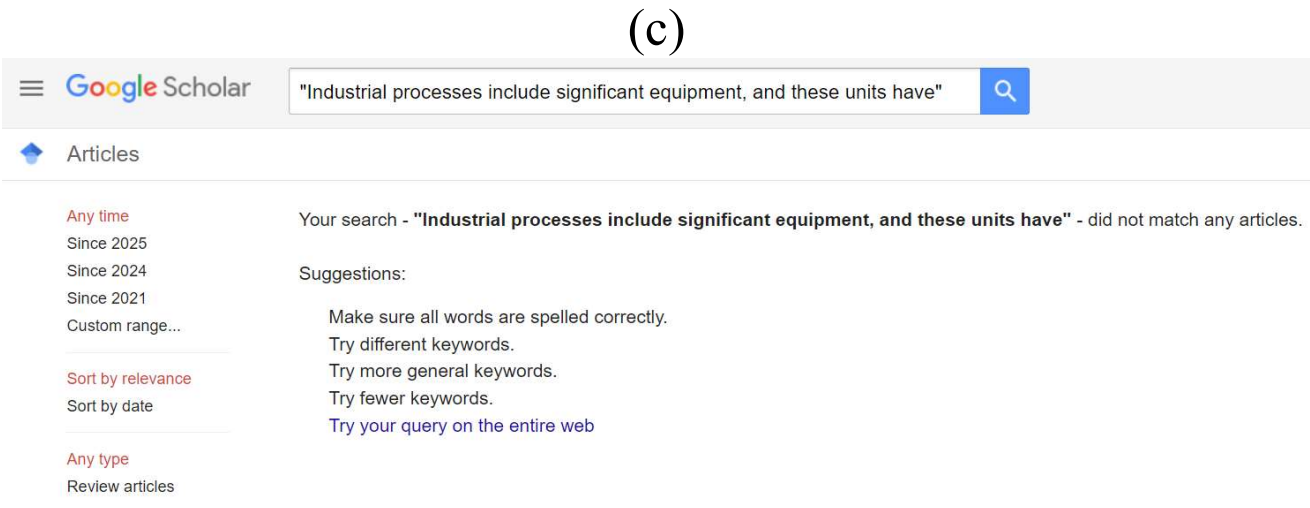

(d)

Fig. 9 Indexing disrupted by the paragraph published in *Journal of Industrial Information Integration* [9].

## C. Normal Indexing example

To clarify how proper indexing should work, we first use one article published in *Applied Energy* recently as an example, which is online after proof without pre-proof policy.

Reference: [10]

Doi: https://doi.org/10.1016/j.apenergy.2025.125526

It can be seen in Fig. 10 that using any words in the article can search this article in Google Scholar.

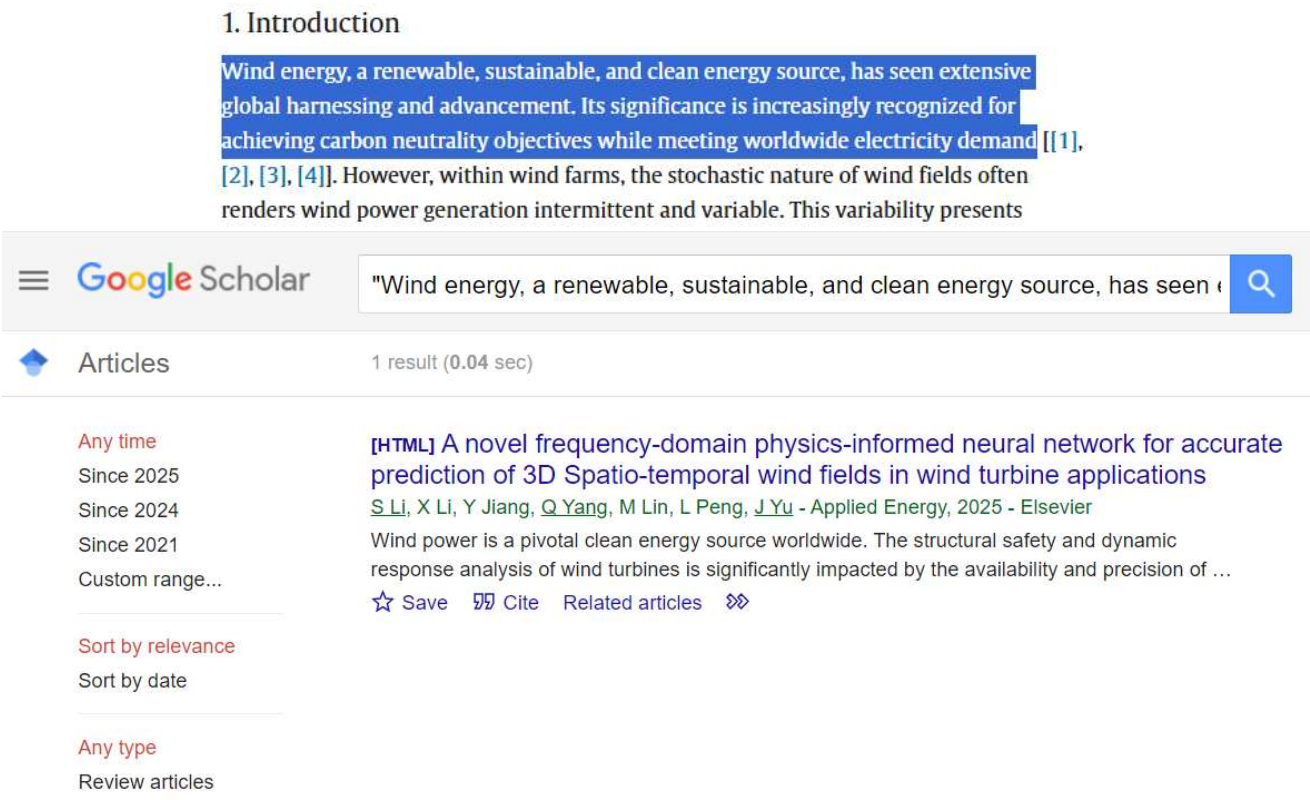

(a)

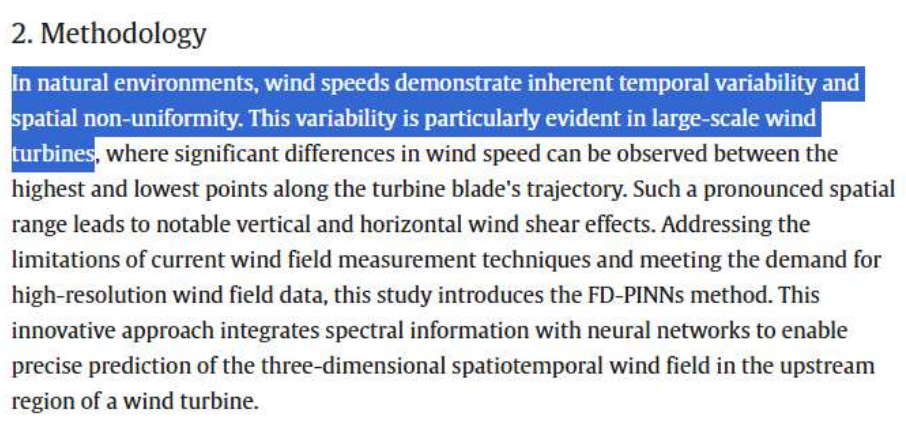

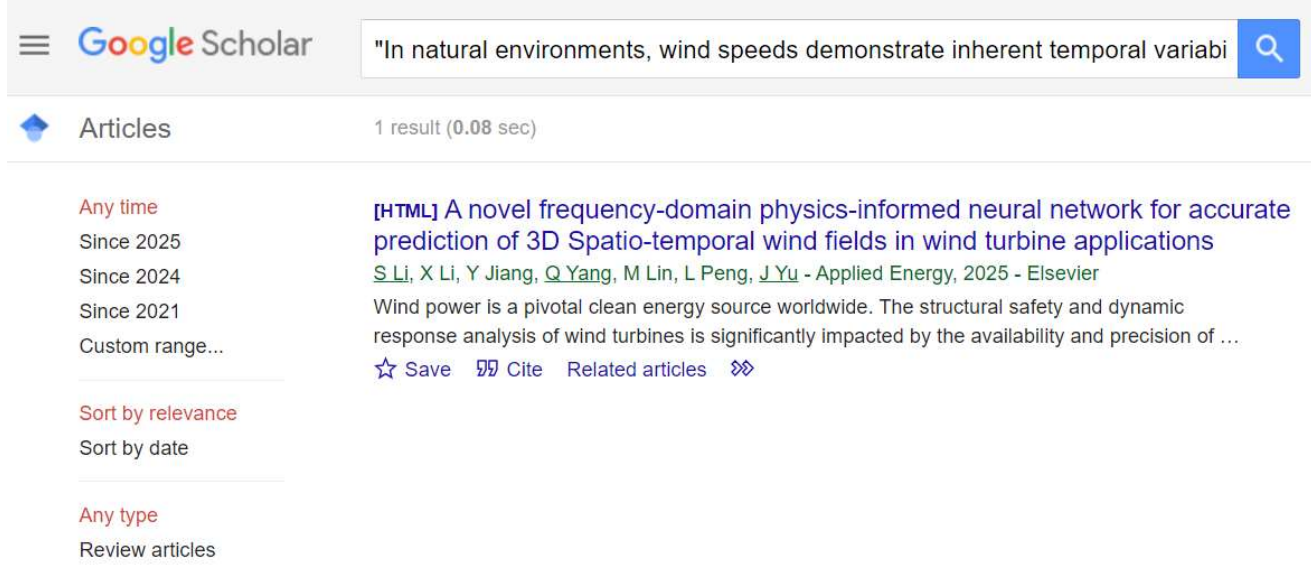

(b)

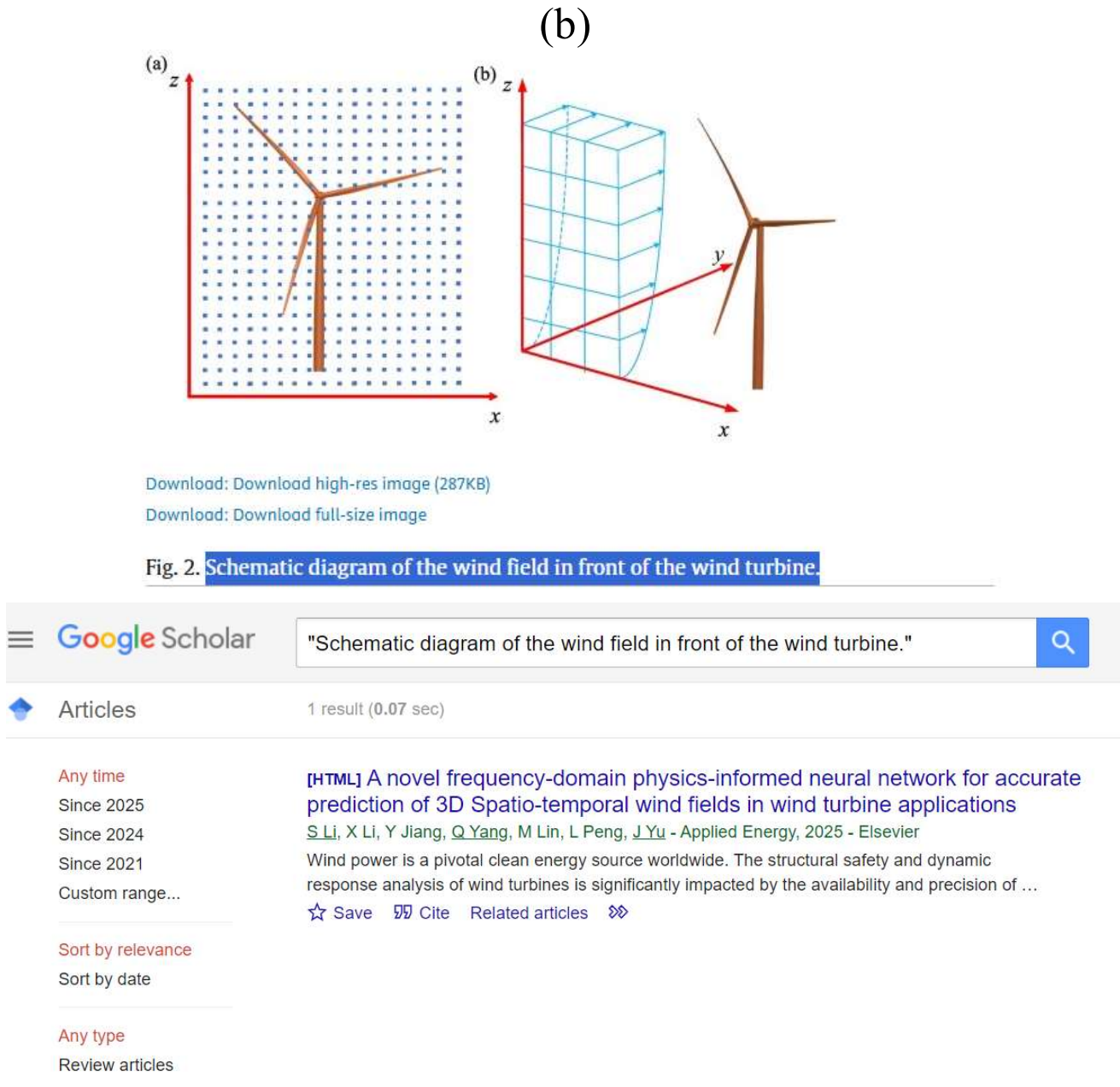

(c)

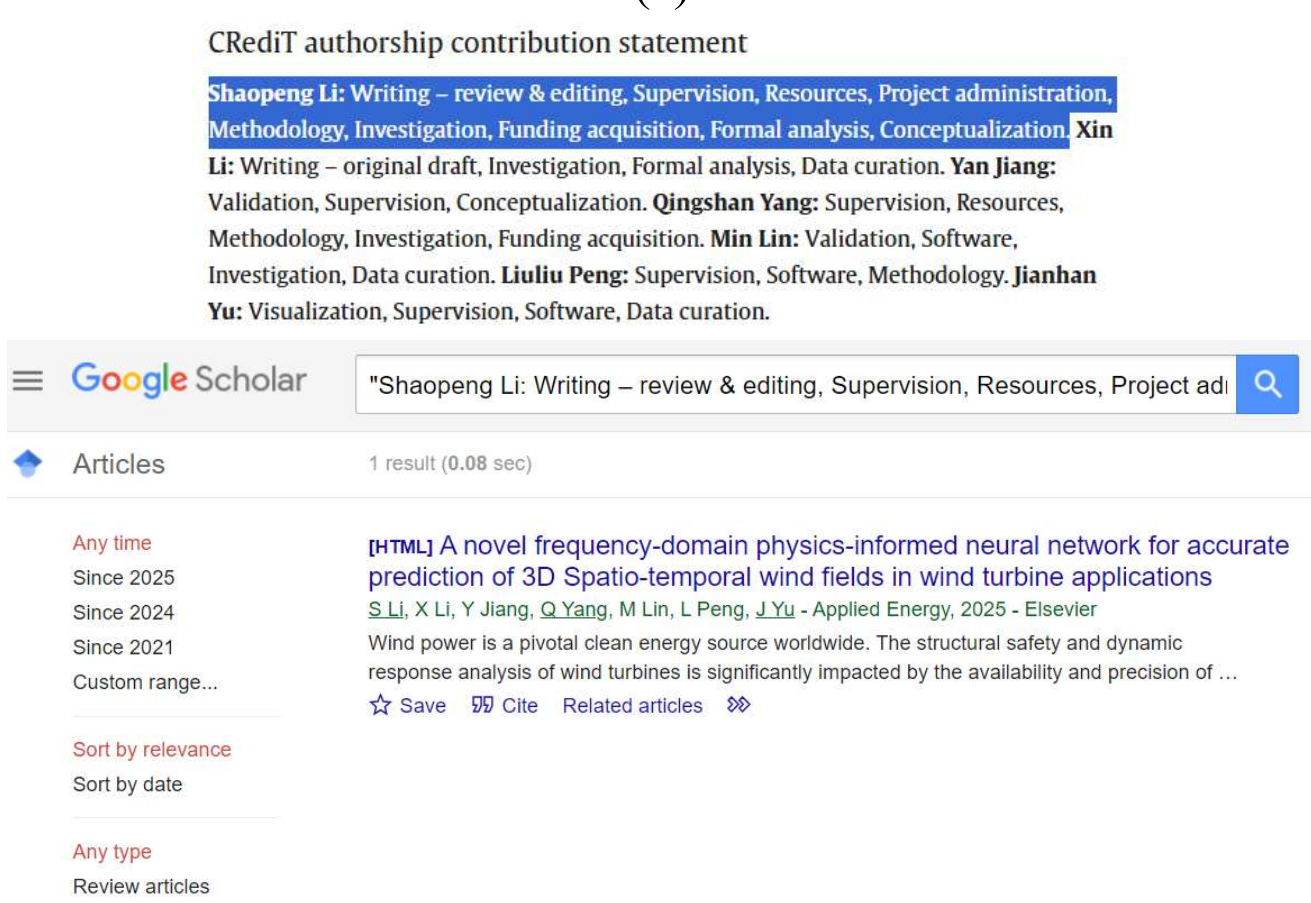

(d)

Fig. 10 Normal indexing published in *Applied Energy* [10].

# III. Harmful Influences

## A. Influence on research availability

For papers that only the titles and abstracts are indexed by Google Scholar [1-5], their research availability is highly limited. Take [4] as an example. This paper focuses on accelerated degradation testing (ADT) modeling using fractional Brownian motion (FBM) considering unit-to-unit variability. However, if we search for articles using the terms "ADT," "FBM," and "unit-to-unit variability" in Google

Scholar, this paper cannot be found, as shown in Fig. 11. This is because the authors did not use these abbreviations in the title or abstract. Although “FBM” appears 30 times and “ADT” 78 times in the main text, the lack of indexing significantly reduces its research availability.

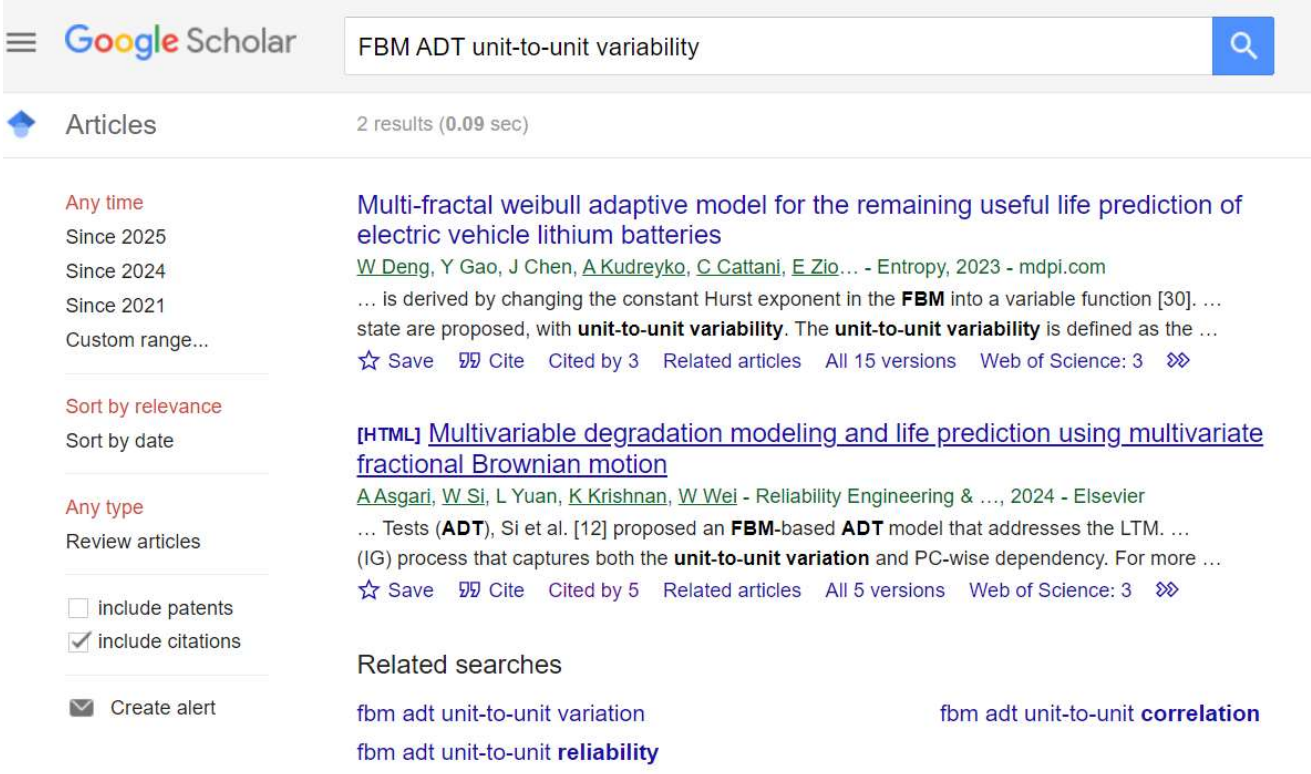

Fig. 11 Search results of given terms in Google Scholar.

## B. Influence on code availability

According to the study in [11], only 36.36% of papers that included GitHub code in the “Data Availability” section published by Elsevier are discoverable via Google Scholar. This limited visibility is largely due to the pre-proof policies adopted by many Elsevier journals. Due to the random disruption discussed in Section II.B, code placed at the end of a paper is less likely to be indexed, significantly reducing the code availability in Elsevier publications.

## C. Influence on research discoverability

Articles published in Elsevier journals with pre-proof policies may have access restrictions, whereas those published elsewhere are fully indexed. For a specific keyword, a search within a restricted article might yield only three corresponding terms, while the same article published elsewhere could provide thirty. Due to the way Google Scholar ranks results, journals without pre-proof policies often appear before those that do, making them more accessible to researchers. Since most users tend to cite the first few search results, this ranking advantage amplifies the visibility and influence of these unrestricted journals. Over time, this effect accumulates, leading to significantly lower citation counts for Elsevier journals with pre-proof policies. However, this discrepancy does not necessarily reflect differences in quality but rather the impact of search engine algorithms, which introduce a systematic bias in favor of certain publications.

# IV. Conclusions

This study investigated the impact of Elsevier’s Pre-proof Policy on Google Scholar indexing. Based on the studies, several conclusions can be drawn:

- Due to pre-proof publication policies of Elsevier, content in the articles published by these journals are not indexed by Google Scholar, including introductions, methods, results, discussions, conclusions, appendices, and data availability statements.
- This index problem has harmful impacts on research availability, code availability and research discoverability.

To solve this problem, both Elsevier and Google Scholar must modify publishing policies or enhance indexing practices. Additionally, before their modifications, authors can use the following strategies to mitigate the issue and ensure broader discoverability of their research:

1. Try to ask for making the paper online after proof.
2. Don’t make any changes in the proof stage.

The second strategy, in particular, is supported by [12]. Although this paper was published in a journal that adopts a pre-proof policy, it does not encounter indexing problems, as shown in Fig. 12.

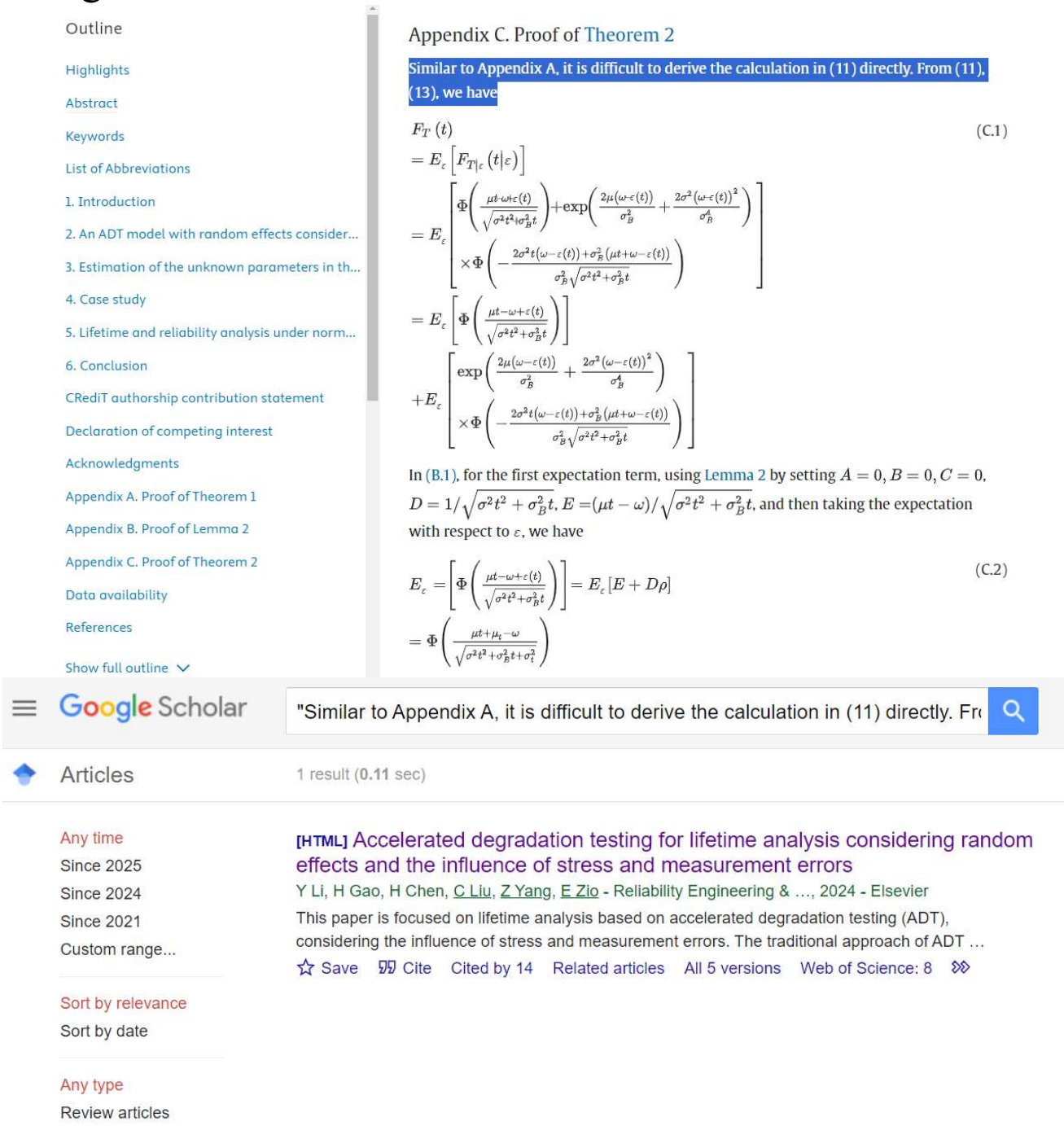

Fig. 12 Normal indexing published in RESS [12].